\documentclass[11pt]{article}
\usepackage{hyperref}
\usepackage{amssymb}
\usepackage{graphicx}
\usepackage{subcaption}
\usepackage{float}
\usepackage{slashed}
\usepackage{amsmath}
\usepackage{amssymb}
\usepackage{tikz}
\usepackage{fixmath}
\usepackage{tabularx}
\usepackage{breqn}
\usepackage{authblk}
\usepackage{array}
\usepackage{dsfont}
\usepackage{relsize}
\usepackage{cite}
\usepackage{physics}
\usepackage[section]{placeins}
\usepackage[a4paper, total={6.8in, 10in}]{geometry}
\numberwithin{equation}{section}
\newcommand{\be}{\begin{equation}}
\newcommand{\ee}{\end{equation}}
\newcommand{\bea}{\begin{eqnarray}}
\newcommand{\eea}{\end{eqnarray}}
\newcommand{\ba}{\begin{aligned}}
\newcommand{\ea}{\end{aligned}}

\begin{document}
\title{Vacuum fluctuations in Rainbow space-time: Study of Casimir effect}

\author[1]{Bhagya. R \thanks{22phph03@uohyd.ac.in, r.bhagya1999@gmail.com}}
\affil[1]{School of Physics, University of Hyderabad, Central University P.O, Hyderabad-500046, Telangana, India
}

\author[2,3]{Suman Kumar Panja \thanks{sumanpanja@ustc.edu.cn, sumanpanja19@gmail.com
}}

\affil[2]{Wilczek Quantum Center, Shanghai Institute for Advanced Studies USTC, Shanghai 201315, China}

\affil[3]{University of Science and Technology of China, Hefei 230026, China}

\date{}
\maketitle

\begin{abstract}

We investigate the Casimir effect in rainbow space-time, focusing on leading-order corrections to the Casimir energy and force. Starting with the scalar field Lagrangian in rainbow space-time, with parallel plates introduced through $\delta$-function potentials, we find the corresponding energy-momentum tensor. We obtain the vacuum expectation value of this energy-momentum tensor by expressing it as a quadratic operator acting on the  Green's function. By solving the Euler-Lagrange equation of a scalar field in rainbow space-time, we obtain the Green's function solutions. Employing these Green’s function solutions in the vacuum expectation value of the energy-momentum tensor, we obtain the modified Casimir energy and Casimir force expressions in rainbow space-time. We study the variation of the deformed Casimir force and energy with the distance between the plates for different choices of rainbow functions. Our results show that for two choices of rainbow functions, the absolute value of the Casimir energy and force is decreasing or increasing, whereas for one specific choice of rainbow functions, it remains the same as the standard result in Minkowski space-time. Comparing our result with experimentally measured value of Casimir force, we obtain the bound on the rainbow parameter dependent terms to be of the order of ${10}^{-24}$.

\end{abstract}

\section{Introduction}\label{sec-1}

Unification of gravity with quantum theory is one of the most challenging areas of research in modern physics \cite{roveli}. A consistent theory of quantum gravity requires the fusion of quantum mechanics with general relativity, and several approaches have been proposed in this direction. Loop quantum gravity, string theory, non-commutative geometry, etc., are some of those approaches to describe the quantum nature of space-time \cite{roveli, QG, string, connes, seiberg, doplicher}. A common feature of all these approaches is the existence of a fundamental length scale that marks the threshold below which the quantum nature of space-time becomes evident. 

 Rainbow gravity is one such approach, which is the curved-space generalisation of doubly special relativity (DSR), that emerges by  introducing two fundamental constants: the speed of light and the fundamental length scale \cite{GAC1, Gilkman, lee2}.  The rainbow space-time (RST) is based on the idea that the geometry of space-time depends on the energy of the particle probing the space-time.   Here, the space-time metric is energy-dependent, leading to a modified dispersion relation. In recent times, the RST has gained considerable attention, and several works have been done in this area \cite{yi,CEM, SBRG2, SBRG1, EH}. The thermodynamics of modified black hole solutions in RST was studied in \cite{yi}, where it was found that the temperature and entropy are modified by the modified dispersion relation. The anisotropic neutron stars in Rastall-Rainbow gravity background were investigated in \cite{CEM}. The author has considered both Rastall-Rainbow gravity and anisotropy simultaneously and obtained the values for mass and radius, which support astrophysical systems
such as the LMXB NGC 6397, and showed that the hyperon puzzle problem can be solved in this way. In \cite{SBRG2}, a relativistic ideal gas was considered, and its thermodynamic properties were studied in the background of $\kappa$-deformed space-time and Rainbow gravity. This gave rise to modifications in both the Stefan-Boltzmann constant and the Debye theory of specific heat. The Abelian and non-Abelian gauge theories in RST were constructed in \cite{EH}. Employing the modified field strength tensor, it was shown that the time-dependent Aharonov-Bohm effect is, in general, non-zero in RST. Maxwell's theory in RST was used to study modifications to the Stefan-Boltzmann law and Wein's displacement law for three choices of rainbow functions in \cite{SBRG1}. The scalar theory in RST was studied in \cite{lee, joao1}, where the scalar field Lagrangian in the RST framework was obtained, and it was shown that the plane wave solution is also a solution in RST.  The effect of gravity at higher energy scales on spinless particles and fermions in different space-times in RST was investigated in \cite{zare, bake1}. 
 
Since quantum gravity theories predict the existence of a minimal length scale, below which the quantum effects of space-time predominate, the existence and significance of a minimal length in physical phenomena is of immense interest.  The Casimir effect is one such phenomenon in which the length scale plays an important role in the associated effects \cite{HC}. The studies in this direction showed that the separation between the plates is of the order of micrometre or less  \cite{HC,casimir1,kam}. Here, we investigate the role played by the minimal length scale introduced by the energy dependence of space-time in the Casimir effect.

Modifications of the Casimir effect arising from variations in space-time dimensionality, the presence of a medium between plates, and thermal effects have been investigated using quantised field theory \cite{kam,Brevik}. The Casimir self-stress and dynamical Casimir effect were discussed in \cite{casimir1}. Experiments were carried out to measure the Casimir force between flat metal plates, and the results obtained were found to agree with the theoretical values \cite{jtep, sparnaay, SKL}. In \cite{SKL}, the experiment was carried out using an electromechanical system and a torsion pendulum, and the results were accurate to within $5$ per cent. In \cite{PRDB63, science}, the presence of the Casimir force in microelectromechanical systems (MEMS) is reported.

Studies on Casimir force and Casimir energy in various non-commutative space-times such as Moyal space time and $\kappa$-deformed space-time were carried out in \cite{moyal1, moyal2, suman1,SumK, suman2, suman3}. In \cite{moyal2}, the Casimir effect between two parallel plates in a non-commutative space-time was investigated using a coherent state approach, and it was realised that boundary corrections are equally important as the non-commutative volume corrections. The modifications to the Casimir effect for a complex scalar theory were investigated in 2+1 dimensions in the Moyal space-time \cite{moyal1}. In \cite{suman1}, the scalar theory in $\kappa$-space-time is considered, and by using the Green's function method, the Casimir force and energy associated with two parallel plates are obtained. The results from this work show that the $\kappa$-deformed Casimir force increases linearly with the deformation parameter. The effect of the extra dimension on the Casimir effect in $4+1$-dimensional DFR space-time is studied in \cite{suman2}. In \cite{suman3}, the fall in the Casimir energy due to a weak gravitational field in $\kappa$-space-time is investigated, and the results show that the equivalence principle is preserved in the quantum gravity regime as well. In \cite{mota}, the modification of the quantum vacuum energy due to the non-trivial topology of the Einstein universe in RST is investigated using Epstein-Hurwitz and Riemann’s zeta functions.

In this study, we investigate modifications to the Casimir energy and force in an energy-dependent space-time, the rainbow space-time (RST), valid up to first order in the rainbow parameters, using the Green's function method. We consider the scalar field in RST, and the two parallel plates are introduced through $\delta$-function potentials. Using the total Lagrangian, which is the sum of the Lagrangian of the scalar field and the interaction potential due to the parallel plates in RST, the expression for the corresponding energy-momentum tensor in RST is obtained. The vacuum expectation value of the energy-momentum tensor at two nearby points is equated with the Green's function. This is done by expressing the vacuum expectation value of the energy-momentum tensor as a quadratic operator acting on the time-ordered product of fields at two nearby points. By solving the equation of motion corresponding to a scalar field interacting with a potential due to the parallel plates in the RST background, we find the Green's function solution in RST. This result, along with the vacuum expectation value of the $00$-component of the energy-momentum tensor, gives the expression for Casimir energy in RST. In our study, we take three choices of rainbow functions. Our results show that for two of the choices of rainbow functions, both the Casimir energy and the Casimir force in RST are modified, whereas for the third choice, they remain the same as in Minkowski space-time. Under the strong interaction limit, we find that for the first choice of rainbow functions, the absolute value of Casimir energy and force decreases, whereas for the third choice, corrections enhance the absolute value of both the energy and force. Further, we study the variation of the Casimir energy and force with plate separation for different choices of rainbow functions, respectively. Using the measured value of Casimir force \cite{exp}, we show that the rainbow parameter dependent term ($a \varepsilon$) should be less than ${10}^{-24}$.

This paper is organised as follows:  A summary about RST is given in section \ref{sec-2}. The important definitions and the three choices of rainbow functions are presented here. The construction of modified Lagrangian and the corresponding energy-momentum tensor in RST are given in section \ref{sec-3}. The vacuum expectation value of the $00$-component of the energy-momentum tensor is discussed in this section. The deformed equations of motion and the Green's function solution are obtained in section \ref{sec-4}. In section \ref{sec-5}, we derive the Casimir force and Casimir energy between parallel plates in RST. The concluding remarks are presented in section \ref{sec-6}.

\section{Rainbow space-time}\label{sec-2}

Rainbow space-time (RST) is an energy-dependent space-time where the dispersion relation, space-time interval, etc., become observer dependent \cite{lee, smolin2}.  The dispersion relation for a particle of mass $m$ in this space-time is
\be\label{dispersion}
E_o^2 f^2\Big(\frac{E_o}{E_P}\Big) - p^2 g^2\Big(\frac{E_o}{E_P}\Big) = m^2.
\ee
Here $f=f(\frac{E_o}{E_P})$ and $g=g(\frac{E_o}{E_P})$ are called rainbow functions, where $E_o$ and $E_P$ are the energy of the test particle/observer and Planck energy, respectively.
The modified line element in RST is \cite{joao1}
\be \label{line element}
ds^2
= -\frac{dt^2}{f^2(\frac{E_o}{E_P})}  + \frac{dx_i^2}{g^2(\frac{E_o}{E_P})}.
\ee
In the limit $f \to 1$ and $g \to 1$, the above equations reduce to the energy-momentum relation and the line element of the Minkowski space-time.
The space-time metric in RST can be written as
\be \label{metric}
\hat{\eta}_{\mu \nu} =\text{diag }\left[-\frac{1}{f^2 (\frac{E_o}{E_P})}, \frac{1}{g^2 (\frac{E_o}{E_P})}, \frac{1}{g^2(\frac{E_o}{E_P})},\frac{1}{g^2(\frac{E_o}{E_P})}\right],
\ee
This modified dispersion relation and the line element remain invariant under the non-linear Lorentz generators $K^i = U^{-1} L_0^i U$. The transformation connecting inertial observers is now implemented by a non-linear realisation of the Lorentz transformation \cite{lee, lee2},
\be \label{energy}
U_{\mu}(p) \equiv U_{\mu} (E_o,p_i) = (U_0,U_i)= \Big(E_o f(\frac{E_o}{E_P}),p_i g(\frac{E_o}{E_P})\Big).
\ee
The corresponding transformation in position space is obtained by imposing the condition that $U_{\mu}(p)$ must contract with  $U^{\mu}(x)$ \cite{joao1}, i.e.,
\be \label{map}
U_{\mu}(p)U^{\mu}(x) = p_{\mu}x^{\mu},
\ee
where $\mu=0,1,2,3$ and the transformation in the position space is,
\be \label{T1}
U^{\mu} (x)= (U^0, U^i)=\Big(\frac{t}{f(\frac{E_o}{E_P})}, \frac{x^i}{g(\frac{E_o}{E_P})}\Big).
\ee
In our study, we are focusing on three choices of rainbow functions \cite{lee2, lee, ellis, Jacob, Amel}. The first choice, which is obtained from the condition that the velocity of light is constant in RST \cite{lee2, lee}, is
\be \label{rf1}
f(E_o) = g(E_o) = \frac{1}{1- \lambda \frac{E_o}{E_P}}.
\ee
The studies on hard spectra from gamma-ray bursts lead to the second choice,
\be \label{rf2}
f(E_o) = \frac{e^{\frac{\beta E_o}{E_{P}}}-1}{\beta \frac{E_o}{E_{P}}},~~~g(E_o) = 1,
\ee
and the investigations in the field of loop quantum gravity and non-commutative space-time suggest the third choice of rainbow functions \cite{Amel},
\be \label{rf3}
f(E_o) = 1 ,~ g(E_o) =\sqrt{1 - \eta \Big(\frac{E_o}{E_P}\Big)^n}.
\ee
Note here that $\lambda$, $\beta$, and $\eta$ are the rainbow parameters corresponding to each choice of the rainbow functions, and in our study, we consider only up to first order in these rainbow parameters.

\section{Modified Lagrangian and Energy-momentum tensor} \label{sec-3}

In this section, to study the Casimir effect between two parallel plates in RST, we construct the Lagrangian as the sum of the scalar field Lagrangian in RST and the interaction Lagrangian due to the parallel plates. Here, we consider the parallel plates to be located at $z=z_1$ and $z=z_2$ along the z-axis and model them using delta-function potentials. Using the modified Lagrangian, we also obtain the expression for the corresponding  deformed energy-momentum tensor. The total energy associated with the parallel plates is related to the vacuum expectation value of the $00$ component of the energy-momentum tensor in RST, which in turn is expressed in terms of the Green’s function $\hat{G}(x,x')$. We commence with the Lagrangian of a scalar particle interacting with a potential $\hat{V}$, in RST, given by
\be \label{Lag}
\hat{L} = -\frac{1}{2} \hat{\eta}^{\mu \nu} \partial_{\mu}\phi(x) \partial_{\nu} \phi(x) - \frac{1}{2} \hat{V} \phi^2(x)
\ee
where $\phi= \phi(x)$ is the field defined in the Minkowski space-time. This is true because the solution to the Klein-Gordon equation in RST is $A e^{-i p_{\mu} x^{\mu}}$, where $x_{\mu}$ and $p_{\mu}$ are the position and momentum in Minkowski space-time \cite{joao1}. Note that in eq.(\ref{Lag}), $\hat{\eta}^{\mu \nu} = \text{diag} (-f^2(E_o), g^2(E_o), g^2(E_o), g^2(E_o))$ is the deformed metric in RST and $f(E_o)$ and $g(E_o)$ are the rainbow functions. Here $\hat{V}$ is the background potential, which introduces the parallel plates kept along the $z$ axis at $z=z_1$ and $z=z_2$ is \cite{casimir1, casimir2}
\be \label{potential}
\hat{V} = \lambda_1 \delta(\hat{z}- \hat{z_1})  +  \lambda_2 \delta(\hat{z}- \hat{z_2}) = |g(E_0)| \Big(\lambda_1  \delta(z-z_1) + \lambda_2 \delta(z-z_2) \Big) =|g(E_0)|V.
\ee
To obtain the above equation, we used eq.(\ref{T1}) and the properties of the delta function. Now employing eq.(\ref{potential}) in eq.(\ref{Lag}),
\bea \nonumber
\hat{L} &=& -\frac{1}{2} \hat{\eta}^{\mu \nu} \partial_{\mu}\phi \partial_{\nu} \phi - \frac{|g(E_o)|}{2} \Big[\lambda_1  \delta(z-z_1) + \lambda_2 \delta(z-z_2) \Big] \phi^2 \\
&=& -\frac{1}{2} \left[- f^2(E_o)(\partial_t\phi)^2 +g^2(E_o) (\partial_i \phi)^2 \right] -\frac{|g(E_0)|}{2} \left[ \lambda_1  \delta(z-z_1) + \lambda_2 \delta(z-z_2) \right] \phi^2
\eea
In our study, we are considering three different choices of rainbow functions (given in eq.(\ref{rf1}), eq.(\ref{rf2}), and eq.(\ref{rf3})), which can be expressed in a general form as
\be \label{functions}
f(E_o) = 1 + \alpha E_o + O(\alpha^2); ~~g(E_o) = 1+kE_o + O(k^2)
\ee
where $\alpha = \frac{\lambda}{E_P}, \frac{\beta}{2E_P}, 0$ and $k= \frac{\lambda}{E_P},0, -\frac{\eta}{2E_P}$, respectively, for the three choices of rainbow functions. Note that we are taking $n=1$ in the third choice of rainbow functions. Using the general form of $f(\frac{E_o}{E_P})$ and $g(\frac{E_o}{E_P})$ given in eq.(\ref{functions}) and considering up to first order in $\alpha$ and $k$, we obtain
\be \label{Lag1}
\hat{L} = -\frac{1}{2} \eta^{\mu \nu} \partial_{\mu} \phi \partial_{\nu} \phi + \Big[ \alpha E_o (\partial_t \phi)^2 - kE_o (\partial_i \phi)^2 \Big]  - \frac{1}{2} (1 + k E_o) \Big[ \lambda_1 \delta(z - z_1)+ \lambda_2 \delta (z-z_2) \Big] \phi^2  
\ee
where $\eta^{\mu \nu } = \text{diag}(-1,+1, +1,+1)$. The equation of motion corresponding to the scalar field interacting with parallel plates in RST is
\be \label{eqm}
-(1+ \alpha E_o)^2 \partial_t^2\phi + (1+kE_o)^2 (\partial_x^2 \phi + \partial_y^2 \phi+ \partial_z^2 \phi )= (1+kE_o) V \phi.
\ee
Using eq.(\ref{Lag1}) in the general expression for the energy-momentum tensor we find \cite{suman3}
\bea
\hat{T}_{\mu \nu} &=&\partial_{\mu} \phi \partial_{\nu} \phi  + \hat{\eta}_{\mu \nu} \hat{L}
\eea
Note that in the Minkowski limit, i.e., when $\alpha, k \to 0$, $\hat{T}_{00}$ goes to the corresponding expression in Minkowski space-time. Using the explicit form of $\hat{L}$ in the above equation and after further simplifications, we find the $00$ component of the modified energy-momentum tensor as
\be \label{T001}
\hat{T}_{00} = \frac{1}{2} (\partial_t \phi)
^2 + \frac{1}{2} (\partial_i \phi)^2  + (kE_o- \alpha E_o) (\partial_i \phi)^2 +\Big( \frac{1}{2} + \frac{kE_o}{2} - \alpha E_o \Big)V \phi^2
\ee
For simplifying further, we use eq.(\ref{eqm}) in the above equation, and taking up to first order in $\alpha$ and $k$ gives
\be
\hat{T}_{00} = \frac{1}{2} (\partial_t \phi)
^2 - \frac{1}{2} \phi \partial_t^2 \phi+ \frac{1}{4} \partial^2_i \phi^2  + \frac{(k -\alpha)}{2} E_o \partial^2_i\phi^2, 
\ee
the $00$-component of the energy-momentum tensor, which represents the energy density. The Casimir energy associated with the parallel plates is obtained by taking the vacuum expectation value of the $00$-component of the energy-momentum tensor. This is done by considering two nearby points $x$ and $x'$ and expressing the energy-momentum tensor as a quadratic derivative operator acting on the vacuum
expectation value of the time-ordered product of fields at these nearby points. i.e., we write
\bea
< \hat{T}_{00}> = \frac{1}{2} \left\{ \bigg(\frac{\partial}{\partial t} \frac{\partial}{\partial t'} - \partial_t^2 \bigg) <\phi(x) \phi(x')> \right\}_{x=x'} + \Bigg( \frac{1}{4} + \frac{kE_o-\alpha E_o}{2}  \Bigg) \left\{ \partial^2_i (<\phi(x) \phi(x')>)_{x=x'} \right\}. \nonumber \\
\eea
Here $\phi(x)$ and $\phi (x')$ are the fields defined in two nearby points $x$ and $x'$, and $<\phi(x) \phi(x')>$ is the vacuum expectation value of the time-ordered product of the fields. Using the relation $<\phi(x) \phi(x')> = \frac{1}{i} \hat{G}(x, x')$ where $\hat{G}(x,x')$ is the Green’s function solution obtained by solving the Euler-Lagrange equation corresponding to the Lagrangian of the scalar field given in eq.(\ref{Lag1}), in the above equation, we get
\bea \label{VE1}
< \hat{T}_{00}> &=& \frac{1}{2i} \left\{ (\frac{\partial}{\partial t} \frac{\partial}{\partial t’} - \partial_t^2) \hat{G}(x, x')\right\}_{x=x'} + \frac{1}{i} \Bigg( \frac{1}{4} + \frac{k E_o-\alpha E_o}{2}  \Bigg) \left\{ \partial^2_i ( \hat{G}(x, x)) \right\}
\eea
In our study, the plates are kept in the $z$ direction. Thus, the Fourier transformation of Green’s function is expressed as
\be \label{GF1}
\hat{G}(x, x') = \int \frac{d \omega}{2 \pi} \frac{d^2 k_{\perp}}{(2\pi)^2} e^{-i\omega (t-t’)} e^{ik_{\perp} (x_{\perp }- x'_{\perp})} \hat{g} (z,z')
\ee
where $\hat{g}(z,z')$ is the reduced Green’s function. Using eq.(\ref{GF1}) in eq.(\ref{VE1}) gives
\be
< \hat{T}_{00} >
=\int \frac{d\omega}{2\pi}\frac{d^2k_{\perp}}{(2\pi)^2}
\left\{
\frac{1}{2i}\,2\omega^2\hat g(z,z)
+\frac{1}{i}
\Bigg(
\frac{1}{4}+\frac{kE_o -\alpha E_o}{2}
\Bigg)
\partial_i^2 \hat g(z,z)
\right\}.
\ee
Using Gauss-divergence theorem, the second term (inside ${ }$) in the above equation goes to zero, and the above equation becomes
\bea \label{VE2}
< \hat{T}_{00}> &=& \int \frac{d \omega}{2 \pi} \frac{d^2 k_{\perp}}{(2\pi)^2} \left\{ \frac{1}{2i} 2\omega^2 \hat{g} (z,z) \right\}
\eea
For evaluating the above integral, we need to find the reduced Green’s function solutions in RST.

\section{Deformed Equations of motion and Green’s function solutions} \label{sec-4}

In this section, we obtain the Green’s function solution by solving the equation of motion corresponding to a scalar field interacting with the background potential due to two parallel plates kept at $z=z_1$ and $z=z_2$ in RST. We commence with the Lagrangian for the scalar particle interacting with a potential $\hat{V}$ in RST given in eq.(\ref{Lag1}) and find the corresponding equation of motion, valid up to first order in $\alpha$ and $k$, as
\be
\Big( -\partial_t^2 + \partial_x^2 + \partial_y^2 + \partial_z^2 -V -2\alpha E_o \partial_t^2 + 2kE_o (  \partial_x^2 + \partial_y^2 + \partial_z^2) -kE_o V \Big) \phi(x)=0
\ee
The corresponding Green’s function solution is
\be \label{GE1}
-\Big( -\partial_t^2 + \partial_x^2 + \partial_y^2 + \partial_z^2 -V -2\alpha E_o \partial_t^2 + 2kE_o (
\partial_x^2 + \partial_y^2 + \partial_z^2) -kE_o V \Big)\hat{G}(x,x')=\frac{\delta^4 (x-x')}{\sqrt{-g}},
\ee
where $\sqrt{-g} = 1- (3k +\alpha)E_o$. We solve the above equation perturbatively assuming the solution $\hat{G}(x,x')$ to be of the form
\be \label{Ghat}
\hat{G}(x,x') = G(x,x') +(\alpha + k )E_o G_1 (x,x'),
\ee
where $G(x,x')$ is the Green’s function solution in the Minkowski space-time.
Using this in eq.(\ref{GE1}) we obtain the following equations
\be \label{Ghat1}
-\Big( -\partial_t^2 + \partial_x^2 + \partial_y^2 + \partial_z^2 -V \Big)G(x,x') = \delta^4 (x-x').
\ee
and
\bea \nonumber
-\Big( -\partial_t^2 + \partial_x^2 + \partial_y^2 + \partial_z^2 -V \Big)G_1 (x,x') &=& \frac{(3k+ \alpha)}{(\alpha + k)} \delta^4 (x-x')- \frac{2\alpha}{(\alpha + k)} \partial_t^2 G(x,x') \\&& + \frac{2k}{(\alpha + k)} \Big(\partial_x^2 + \partial_y^2 + \partial_z^2 \Big)G(x,x') -\frac{k}{(\alpha + k)} V G(x,x') \nonumber \\
\eea
We use eq.(\ref{Ghat1}) in the above equation and simplify as
\be \label{Ghat2}
-\Big( -\partial_t^2 + \partial_x^2 + \partial_y^2 + \partial_z^2 -V \Big)G_1(x,x') = \delta^4 (x-x')-2 \frac{(\alpha -k)}{(\alpha +k)} \partial_t^2 G(x,x') + \frac{k}{(\alpha + k)} V G(x,x').
\ee
Using fourier transformation of the Green’s function $ G( x, x')=  \mathlarger{\int}  \dfrac{d \omega}{2 \pi} \dfrac{d^2 k_{\perp}}{(2\pi)^2}  e^{-i\omega (t-t’)} e^{ik_{\perp} (x_{\perp} - {x'}_{\perp})} g(z,z')$ in the eq.(\ref{Ghat1}) and find
\be
-\Big( \partial_z^2+\omega^2-k_{\perp}^2 -V \Big)g(z,z') = \delta(z-z'),
\ee
where $g(z,z')$, the reduced Green’s function solution (under the rotation, i.e., $\omega \to i \zeta$) corresponding to two parallel plates kept at $z=z_1$ and $z=z_2$ in Minkowski space-time is given by \cite{casimir1,casimir2} 
\bea
g(z,z^{\prime})&=& \Bigg[\frac{1}{2\bar{k}}e^{-\bar{k}|z-z^{\prime}|}-\frac{1}{2\bar{k}\Delta}e^{\bar{k}(z+z^{\prime}-2z_{1})}\bigg\{\frac{\lambda_{1}}{2\bar{k}}\Big(1+\frac{\lambda_{2}}{2\bar{k}}\Big)e^{2\bar{k}L} + \frac{\lambda_{2}}{2\bar{k}}\Big(1-\frac{\lambda_{1}}{2\bar{k}}\Big)\bigg\}  \Bigg], where~~\left\lbrace z, z^{\prime}\right\rbrace <z_{1}. \nonumber \\ \label{A11} \\
&=& \Bigg[\frac{1}{2\bar{k}}e^{-\bar{k}|z-z^{\prime}|}-\frac{1}{2\bar{k}\Delta}\bigg\{\frac{\lambda_{1}}{2\bar{k}}\Big(1+\frac{\lambda_{2}}{2\bar{k}}\Big)e^{-\bar{k}(z+z^{\prime}-2z_{2})} + \frac{\lambda_{2}}{2\bar{k}}\Big(1+\frac{\lambda_{1}}{2\bar{k}}\Big)e^{\bar{k}(z+z^{\prime}-2z_{1})}\nonumber \\
&&~~ -\frac{\lambda_{1}\lambda_{2}}{4\bar{k}^{2}}2cosh\bar{k}(z-z^{\prime})\bigg\}  \Bigg], ~~~~~~~~~~~~~~~~~~~~~~~~~~~~~~~~~~~~~~~~~~~~~~~~~where~~z_{1}<\left\lbrace z, z^{\prime}\right\rbrace <z_{2}. \nonumber \\ \label{A12} \\
&=& \Bigg[\frac{1}{2\bar{k}}e^{-\bar{k}|z-z^{\prime}|}-\frac{1}{2\bar{k}\Delta}e^{-\bar{k}(z+z^{\prime}-2z_{2})}\bigg\{\frac{\lambda_{1}}{2\bar{k}}\Big(1-\frac{\lambda_{2}}{2\bar{k}}\Big) + \frac{\lambda_{2}}{2\bar{k}}\Big(1+\frac{\lambda_{1}}{2\bar{k}}\Big)e^{2\bar{k}L}\bigg\}  \Bigg],~where~\left\lbrace z, z^{\prime}\right\rbrace > z_{2}. \nonumber  \\ \label{A13}
\eea
In the above expressions $\bar{k}=\sqrt{k_{\perp}^{2}+\zeta^{2}}$, $L=z_{2}-z_{1}$ and
\be
\Delta=\Big(1+\frac{\lambda_{1}}{2\bar{k}}\Big)\Big(1+\frac{\lambda_{2}}{2\bar{k}}\Big)e^{2\bar{k}L}-\frac{\lambda_{1} \lambda_{2}}{4\bar{k}^{2}}. \label{A14}
\ee
Next, similar to the previous case, we use Fourier transformations of $G_{1}(x,x')$ and $G(x,x')$ in the eq.(\ref{Ghat2}) and find
\be \label{Ghat3}
-\Big( \partial_z^2+\omega^2-k_{\perp}^2 -V \Big)g_{1}(z,z')  = \delta(z-z')+\left\{2\omega^2 \frac{(\alpha -k)}{(\alpha +k)} + \frac{k}{(\alpha + k)} V\right\} g(z,z').
\ee
We solve the above (by following the same calculational procedure given in \cite{suman3}), and we find the reduced Green’s function solution to be
\be
g_1 (z,z')= g(z,z') + \int d\bar{z}g(z,\bar{z}) \Big\{2 \frac{(\alpha -k)}{(\alpha +k)} \omega^2 + \frac{k}{(\alpha + k)} V \Big\} g(\bar{z}, z')
\ee
In the above, $g(z,\bar{z})$ and $g(\bar{z}, z')$ can be obtained from eq.(\ref{A11}), eq.(\ref{A12}) and eq.(\ref{A13}). Here, the contribution from the second term is zero, as the integration has a range either from $z \to z'$ for $z' >z$ or $z' \to z$ when $z>z'$ i.e.,
\be
g_{1} (z,z')= g(z,z')
\ee
and using the above, from eq.(\ref{Ghat}), the complete expression of the reduced Green’s function solution comes to be
\be \label{RGF}
\hat{g}(z,z') = \Big\{1 + (\alpha + k)E_o \Big\} g(z,z'),
\ee
where $g(z,z')$, the reduced Green’s function solutions corresponding to two parallel plates are given in eq.(\ref{A11}) to eq.(\ref{A13}).

\section{Casimir energy and Casimir force between parallel plates in rainbow space-time} \label{sec-5}

In this section, we use the results obtained in the previous section and obtain the expression for Casimir energy and Casimir force between parallel plates in rainbow space-time. The expression for total energy associated with the parallel plates is
\be
\hat{E}_{tot} = \int d^3x \sqrt{-g} <\hat{T}_{00}> .
\ee
Employing eq.(\ref{metric}) and eq.(\ref{VE2}), we find the above equation as
\be
\frac{\hat{E}_{tot}}{A} = \int dz \Big(1- (3k +\alpha)E_o \Big) \frac{d \omega}{2 \pi} \frac{d^2 k_{\perp}}{(2\pi)^2} \left\{ \frac{1}{2i} 2\omega^2 \hat{g} (z,z)  \right\}.
\ee
Under the rotation, $\omega \to i \zeta$, i.e., when we switch to imaginary frequencies, the above equation becomes,
\be \label{energy}
\frac{\hat{E}_{tot}}{A} =- \Big(1- (3k +\alpha)E_o \Big) \left\{ \frac{1}{2} \int  \frac{ d \zeta}{2 \pi} \frac{d^2 k_{\perp}}{(2\pi)^2}  2\zeta^2 \int_{-\infty}^{\infty} dz \hat{g} (z,z) \right\}.
\ee
Note that here we are considering only upto first order in $\alpha$ and $k$, where $\alpha = \frac{\lambda}{E_P}, \frac{\beta}{2E_P}, 0$ and $k= \frac{\lambda}{E_P}, 0,  -\frac{\eta}{2E_P}$, respectively, for the three choices of rainbow functions. Using eq.(\ref{RGF}) in the above equation, we obtain the modified total energy per unit area for the parallel plates to be
\bea  \label{TEPA}
\frac{\hat{E}_{tot}}{A} = \Big\{1-2kE_o \Big\} \Bigg\{  -\frac{1}{2} \int  \frac{ d \zeta}{2 \pi} \frac{d^2 k_{\perp}}{(2\pi)^2}  2\zeta^2 \int_{-\infty}^{\infty} dz g(z,z) \Bigg\}
\eea
Note that in the above equation, the rainbow parameter-dependent modification is coming as an overall multiplicative factor. Using Green’s function expression for the three regions given in eq.(\ref{A11}) to eq.(\ref{A13}) in eq.(\ref{TEPA}), we compute
\bea
\frac{\hat{E}_{tot}}{A} &=& (1-2kE_o ) \Bigg\{  -\frac{1}{2} \int  \frac{ d \zeta}{2 \pi} \frac{d^2 k_{\perp}}{(2\pi)^2}  \Bigg( \dfrac{\zeta^2}{\bar{k}^2} \int_{-\infty}^{\infty} \bar{k} dz \nonumber \\ & & - \dfrac{\zeta^2}{\bar{k}^2} \dfrac{1}{\Delta} \left[ \frac{\lambda_1}{2\bar{k}}
+\frac{\lambda_2}{2\bar{k}}
+\frac{2\lambda_1}{2\bar{k}}\frac{\lambda_2}{2\bar{k}}( 1-e^{-2\bar{k}L} ) - 2\bar{k} L \frac{\lambda_1}{2\bar{k}}
\frac{\lambda_2}{2\bar{k}} e^{-2\bar{k}L} \right]  \Bigg) \Bigg\}
\eea
Using spherical polar coordinates, the above equation can be rewritten as
\bea \nonumber
\frac{\hat{E}_{tot}}{A} &=& (1-2kE_o) \Bigg\{  -\frac{1}{12 \pi^2} \int_0^{\infty}  \bar{k}^3 d\bar{k}   \int_{-\infty}^{\infty}  dz \\ && + \frac{1}{12 \pi^2}  \int_0^{\infty} d\bar{k} \dfrac{\bar{k}^2}{\Delta} \left[ \frac{\lambda_1}{2\bar{k}}
+\frac{\lambda_2}{2\bar{k}}
+\frac{2\lambda_1}{2\bar{k}}\frac{\lambda_2}{2\bar{k}}( 1-e^{-2\bar{k}L} ) - 2\bar{k} L \frac{\lambda_1}{2\bar{k}}
\frac{\lambda_2}{2\bar{k}} e^{-2\bar{k}L} \right]   \Bigg\}
\eea
In the above the first term is a divergent, which we define as bulk energy per unit area of the system. Considering the second term in the above equation and taking $2\bar{k}L =y$, we find
\begin{align} \nonumber \label{totalenergy}
\frac{\hat{\bar{E}}_{tot}}{A}  &= (1-2kE_o ) \left\{ \frac{1}{96 \pi^2 L^3} \int_0^{\infty}  y^2 dy  \frac{\left[ (\frac{y}{L \lambda_1} + \frac{y}{L\lambda_2})e^y + 2(e^y -1-\frac{y}{2}) \right]}{\left[ (1+ \frac{y}{L\lambda_1})(1+\frac{y}{L \lambda_2})e^y -1 \right]}\right\} \\
&= (1-2kE_o ) \left\{ \frac{E_{z_1}}{A}+ \frac{E_{z_2}}{A} + \frac{E_{Cas}}{A}\right\},
\end{align}
where
\be
\frac{E_{z_1}}{A} = \frac{1}{96\pi^2 L^3} \int_0^{\infty} dy \frac{y^2}{(1+ \frac{y}{\lambda_1 L})}; ~~~~~ \frac{E_{z_2}}{A} = \frac{1}{96\pi^2 L^3} \int_0^{\infty} dy \frac{y^2}{(1+ \frac{y}{\lambda_2 L})}
\ee
and
\be
\frac{E_{Cas}}{A} = -\frac{1}{96\pi^2 L^3} \int_0^{\infty} dy \, y^3  \left\{ \frac{ 1 + \frac{1}{y + L \lambda_1} + \frac{1}{y + L \lambda_2}  }{(1+ \frac{y}{L\lambda_1})(1+\frac{y}{L \lambda_2})e^y -1} \right\},
\ee
where $\frac{E_{z_1}}{A} $ and $\frac{E_{z_2}}{A} $ are the self energies of the two parallel plates and $\frac{E_{Cas}}{A}$ is the Casimir energy between two parallel plates per unit area.
Note that the integrals in the above equation are the same as those in the Minkowski space-time \cite{casimir1}. For the first choice of rainbow function, where $k=\frac{\lambda}{E_P}$, eq.(\ref{totalenergy}) becomes
\be
\frac{\hat{\bar{E}}_{tot}}{A}  = \bigg(1-\frac{2 \lambda E_o}{E_P} \bigg) \left\{ \frac{E_{z_1}}{A}+ \frac{E_{z_2}}{A} + \frac{E_{Cas}}{A}\right\}.
\ee
For the second choice, as $k=0$, there is no change in eq.(\ref{totalenergy}). That is, the energy is not modified in RST for the second choice of rainbow functions. For the third choice of rainbow functions where $k=-\frac{\eta}{2 E_P}$,
\be
\frac{\hat{\bar{E}}_{tot}}{A}  = \bigg(1+ \frac{ \eta E_o}{E_P} \bigg) \left\{ \frac{E_{z_1}}{A}+ \frac{E_{z_2}}{A} + \frac{E_{Cas}}{A}\right\}.
\ee
This implies that the Casimir energy associated with the parallel plates kept at $z=z_1$ and $z=z_2$ gets modified for the first and third choices of rainbow functions as
\be \label{cas1}
\hat{E}_{Cas}  = \bigg(1-\frac{2 \lambda E_o}{E_P} \bigg)\, E_{Cas} ~~\text{and}~~\hat{E}_{Cas}= \bigg(1+ \frac{ \eta E_o}{E_P} \bigg)\, E_{Cas},
\ee
whereas for the second choice of rainbow functions, it remains the same as that in the Minkowski space-time. Note that in the limit $\lambda,\eta \to 0$, the above equations reduce to the corresponding results in commutative space-time.
Considering the special case, i.e., in the strong interaction limit, when $\lambda_1, \lambda_2 \to \infty$, the general expression for Casimir force between parallel plates in RST reduces to
\be
\frac{\hat{E}_{Cas}}{A} = - (1-2kE_o)\frac{1}{96\pi^2 L^3} \int_0^{\infty} dy  \frac{y^3 }{e^y -1} = - (1-2kE_o)\frac{\pi^2}{1440 L^3}
\ee
where we have used $\mathlarger{\int}_0^{\infty} dy  \dfrac{y^3 }{e^y -1} = \zeta(4)\Gamma(4)= \frac{\pi^4}{15}$. In this limit, eq.(\ref{cas1}) becomes
\be
\frac{\hat{E}_{Cas}}{A} = - \Big(1-\frac{2\lambda E_o}{E_P} \Big)\frac{\pi^2}{1440 L^3} ,~~\text{and}~~\frac{\hat{E}_{Cas}}{A} = - \Big(1+\frac{\eta E_o}{E_P} \Big)\frac{\pi^2}{1440 L^3} ,
\ee
respectively. Under the condition when $\lambda_1, \lambda_2 \to \infty$, using $\frac{\hat{F}_{Cas}}{A} =
-\frac{\partial}{\partial L}
\left(
\frac{\hat{E}_{Cas}}{A}
\right)$ we obtain the deformed Casimir force
\be \label{deformedF}
\frac{\hat{F}_{Cas}}{A} = - \Big(1-\frac{2\lambda E_o}{E_P} \Big)\frac{\pi^2}{480 L^4} ,~~\text{and}~~\frac{\hat{F}_{Cas}}{A} = - \Big(1+\frac{\eta E_o}{E_P} \Big)\frac{\pi^2}{480 L^4} ,
\ee
for the first and third choices of rainbow functions, respectively. For the second choice of rainbow functions, the Casimir force is the same as that in the Minkowski space-time \cite{casimir1}. We observe that the Casimir energy and force between parallel plates in RST are modified for the first and third choices of rainbow functions. Here, for the first choice of Rainbow functions, correction decreases the absolute value of standard results, similar to what has been shown in \cite{suman3}. For the third choice, modification enhances the absolute value of the Casimir energy and force.
\begin{figure}[h!] 
\includegraphics[scale=.44]{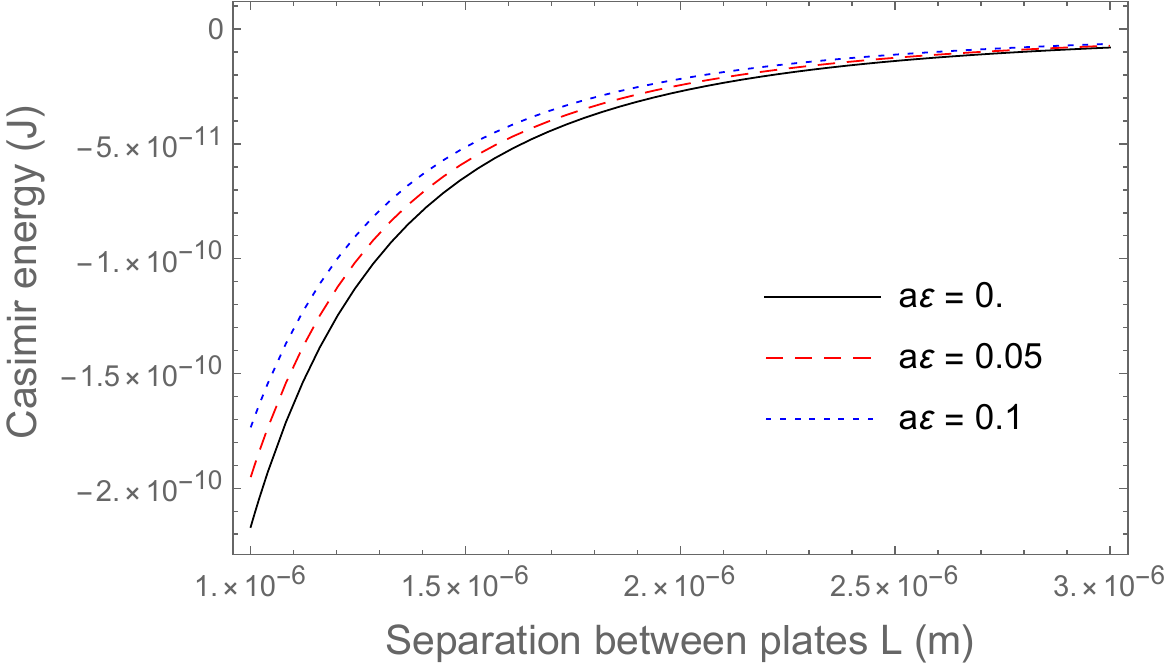}
\includegraphics[scale=.425]{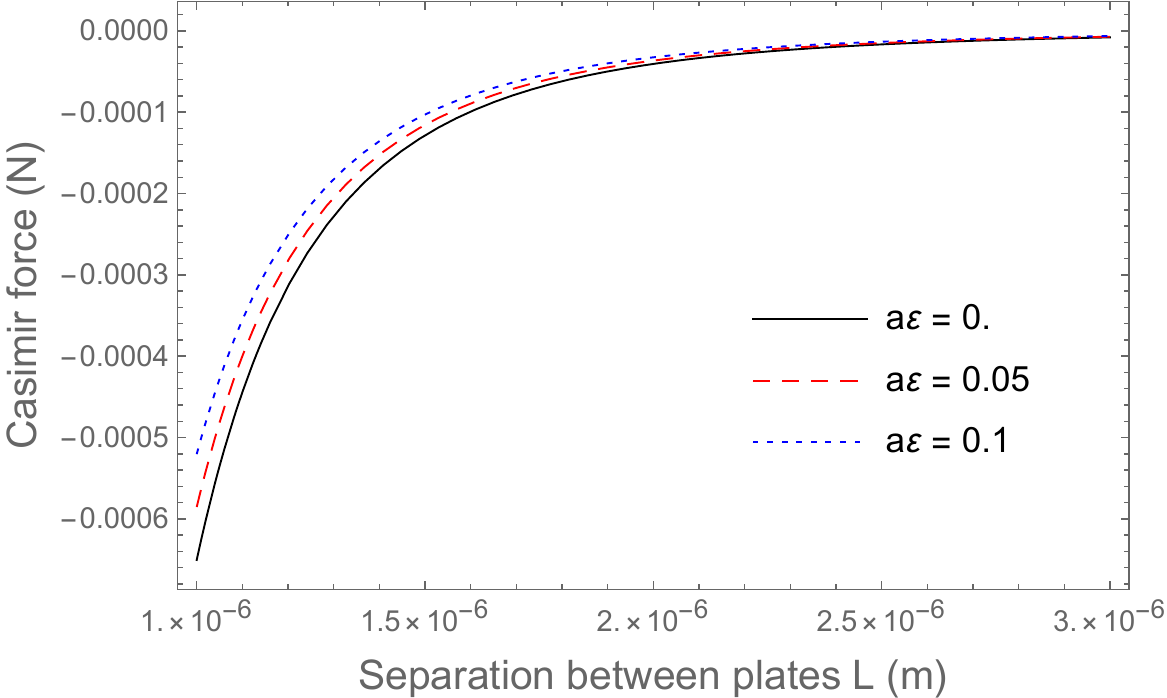}
\caption{Variation of Casimir energy and Casimir force with separation between the plates for first choice of rainbow functions.}
\label{plot1}
\end{figure}
\begin{figure}[h!] 
\includegraphics[scale=.44]{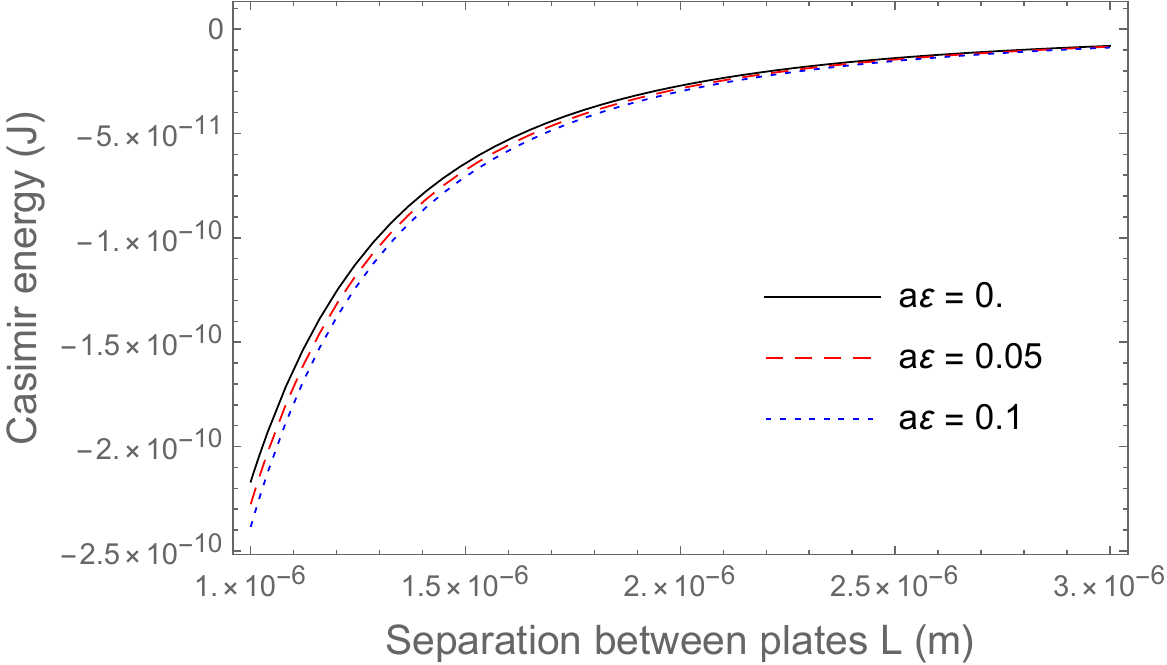}
\includegraphics[scale=.425]{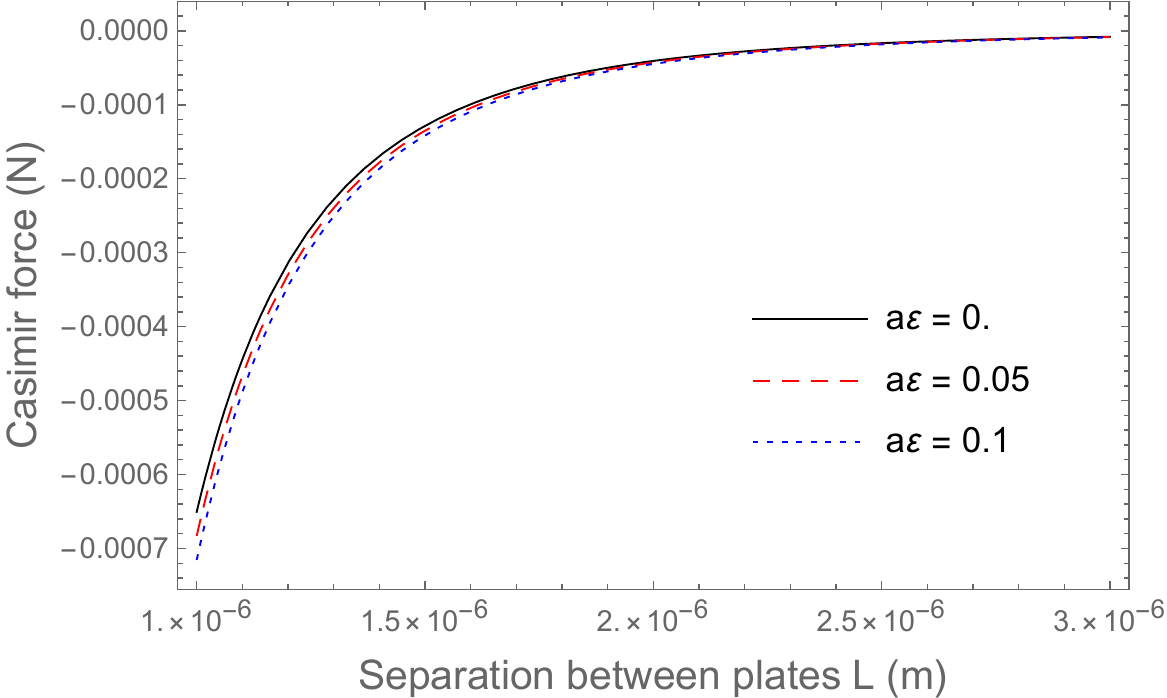}
\caption{Variation of Casimir energy and Casimir force with separation between the plates for third choice of rainbow functions.}
\label{plot3}
\end{figure}
In Fig.\ref{plot1} and Fig.\ref{plot3}, we have investigated the variation of Casimir energy and Casimir force against the separation between the plates for different choices of rainbow functions. In these figures, $a\varepsilon =0$, where $a\varepsilon = \frac{\lambda E_o}{E_P}$ and $\frac{\eta E_o}{E_P}$ for the first and third choices of rainbow functions, respectively, represents the curve corresponding to the standard result \cite{casimir1,kam}. Plots in Fig.\ref{plot1} are obtained for the first choice of rainbow functions, where the Casimir force and energy both decrease as the value of the rainbow parameter dependent term (in the figure, it is $a\varepsilon =\frac{\lambda E_o}{E_P}$) increases. Here, as the rainbow parameter dependence, which is denoted by the value of $a\varepsilon =\frac{\lambda E_o}{E_P}$, increases, curves for both Casimir energy and force in RST background rise from the corresponding curve in Minkowski space-time. For the third choice of rainbow functions also, we have plotted the curve for Casimir energy and Casimir force in RST (given in Fig.\ref{plot3}) with the distance of separation between the plates. Here, the curves corresponding to different values of $a\varepsilon = \frac{\eta E_o}{E_P}$ move away from the Minkowski space-time curve, in the downward direction, as the value of $\frac{\eta E_o}{E_P}$ increases. i.e., as $a \varepsilon$ increases, the absolute value of both Casimir energy and Casimir force increases for the third choice of rainbow functions. Note that the results corresponding to the second choice of rainbow function (where $k=0$) are the same as those in the Minkowski space-time, i.e., there is no modification to the Casimir force and Casimir energy for the second choice of rainbow functions.

For parallel plates separated by a distance of $ 10 \mu m $, the results of measurement of Casimir force gradient in \cite{exp} is $ 8 \cross 10^{-4} Nm^{-2} $ with an error less than 1 percentage. Comparing this with the rainbow dependent correction that we obtained in eq.(\ref{deformedF}), we get a bound on $ a \varepsilon $ ($ = \frac{\lambda E_o}{E_P}$ for first choice of rainbow functions and $ \frac{\eta E_o}{E_P}$ for third choice of rainbow functions, respectively) to be $  a \varepsilon < {10}^{-24} $.

\section{Conclusion} \label{sec-6}

We have studied modifications to the Casimir force and Casimir energy between two parallel plates in RST and examined their variation with plate separation for three different choices of rainbow functions.
We commenced with the scalar field theory in RST, with the interaction of the parallel plates modelled as delta-function potentials. We constructed the deformed energy-momentum tensor in RST from the Lagrangian, and, using the vacuum expectation value of the deformed energy-momentum tensor, calculated the Casimir energy in RST. The vacuum expectation value of the energy-momentum tensor is expressed as a quadratic operator acting on the vacuum expectation value of the time-ordered product of fields at two nearby points. By relating this to the Green's function solution corresponding to a scalar field with an interaction potential due to parallel plates in RST, the modification to the Casimir energy and Casimir force in RST is obtained. Our results showed that for two specific choices of the rainbow functions, the Casimir energy as well as the Casimir force are modified from the standard result obtained in the Minkowski space-time, while for one other choice, it remains the same as that in the Minkowski space-time.

The modification to the Casimir energy, as well as the Casimir force, showed that for two choices of rainbow functions, the absolute value of the Casimir energy is decreasing or increasing, whereas for one specific choice of rainbow functions, it remains the same as that in the Minkowski space-time. We found that for the first choice of rainbow functions, the absolute value of Casimir energy and Casimir force  decreases by  a factor of $\frac{2 \lambda E_o}{E_P}$ while for the third choice it increases by a factor of $\frac{ \eta E_o}{E_P}$. Note that when $\lambda, \eta \to 0$, we get back the corresponding Minkowski results \cite{HC,casimir1,kam}.
Further, for the first and third choices of rainbow functions, by taking different values for $\frac{\lambda E_o}{E_P}$ and $\frac{\eta E_o}{E_P}$, respectively, we studied the variation for Casimir force and Casimir energy with distance of separation in RST in Fig.\ref{plot1} and Fig.\ref{plot3}. Plots in Fig.\ref{plot1} show the variation of Casimir energy and Casimir force for the first choice of rainbow functions. Here, the curves correspond to different values of the rainbow modifications term $a\varepsilon = \frac{\lambda E_o}{E_P}$. The plots showed that as $a\varepsilon$ increases, the curve moves away from the Minkowski space-time curve upward, i.e., the Casimir force decreases. For the third choice of rainbow functions, we considered different values for the rainbow parameters ($a\varepsilon = \frac{\eta E_o}{E_P}$) and plotted the Casimir energy and Casimir force in RST with distance of separation between the plates in Fig.\ref{plot3}. Here, as the rainbow parameter dependent modification increases, the curve moves downwards compared to the curve corresponding to Minkowski space-time. Note that when the rainbow parameters go to zero, we get back the Minkowski space-time results. We have obtained the bound on the value of $a \varepsilon $ by comparing it with the experimental results \cite{exp} and found that $a \varepsilon < {10}^{-24}$.

In \cite{suman1}, the Casimir effect in $\kappa$-deformed space-time is studied in (3+1) dimensions, where the modification to the Casimir force and energy appear as overall multiplicative factors, and the Casimir force is attractive in nature. In our work, also, for two choices of rainbow functions, the RST modification to the Casimir force and Casimir energy is coming as an overall multiplicative factor. For the first choice of rainbow function, the absolute value of the Casimir force decreases with an increase in RST dependence, whereas for the third choice, it increases with an increase in the value of RST correction. The tendency found for the first choice of rainbow functions resembles that of \cite{suman3}, where the correction reduces the quantities from the standard value, whereas in the third choice, where the correction increases the absolute value of the Casimir force and energy is comparable with the results of \cite{suman1}. But this contradicts the results obtained in \cite{AMF} where models based on the Generalised Uncertainty Principle are used to study corrections to the Casimir energy, and the correction varies inversely as the fifth power of the distance of separation between the perfectly conducting parallel plates. The investigation of the Casimir effect in DFR space-time in (4+1) dimensions, carried out in \cite{suman2}, showed that the attractive force between parallel plates increases due to the non-commutativity of space-time, which is consistent with the third choice rainbow functions that we studied. In \cite{CAE}, the Casimir effect is studied for a massive scalar field between parallel plates in a non-commutative space-time. Here, they have used a coordinate-coherent-state approach and plotted the variation of the Casimir force with plate separation for different values of the non-commutative parameter. The results showed that the force is attractive and that the Casimir force in the non-commutative space-time is larger compared to the standard case. This is similar to the results we obtained for the third choice of rainbow functions, where the absolute value of the Casimir force increases with an increase in the rainbow dependence.

The research on the Casimir effect and the progress gained, both theoretical and experimental, shows the importance of investigating the Casimir effect \cite{shen} in various backgrounds. It has expanded its scope to include deep fundamental research and its application to technologies such as atomic force microscopy, micro- and nano-electromechanical systems, and quantum technologies. Our results show that the presence of the energy-dependent space-time leads to a modification to the expression of both the Casimir force and Casimir energy compared to the standard expression, and the nature of this modification, whether it is increasing or decreasing, depends on the choice of rainbow functions. These deformations to the Casimir effect in RST could potentially be probed through comparisons with the experimental results similar to those reported in \cite{Sedmik}.

\section*{Acknowledgements}
We thank E. Harikumar for useful discussions and comments. BR thanks DST-INSPIRE for support through the INSPIRE fellowship (IF220179). SKP would like to acknowledge the support of National Natural Science Foundation of China under Grant No. 12075059.

\section*{Data availability statement}

Data sharing is not applicable to this article as no datasets were generated or analyzed during the current study.

\section*{Conflict of interest }

The authors declare no Conflict of interest.

\end{document}